\documentclass[prd,reprint,amsmath,amssymb,floatfix,aps,longbibliography,superscriptaddress,nofootinbib]{revtex4-2}
\usepackage{graphicx}
\usepackage{dcolumn}
\usepackage{bm}
\usepackage[colorlinks=true,urlcolor=blue,anchorcolor=black,citecolor=blue,filecolor=black,linkcolor=blue,menucolor=blue,linktocpage=true]{hyperref} 
\usepackage{subfigure}
\usepackage{natbib}

\usepackage{booktabs}
\usepackage{colortbl}
\usepackage{collcell}
\usepackage{enumerate}
\usepackage{float}
\usepackage{verbatim}
\usepackage{multirow}
\usepackage{xparse}
\usepackage{tabularx}
\usepackage{cancel}
\usepackage{cleveref}

\usepackage{multirow}

\usepackage[normalem]{ulem}
\usepackage{soul}
\setstcolor{red}

\definecolor{rossos}{cmyk}{0,1,1,0.55}
\definecolor{mygreen}{rgb}{0.27, 0.64, 0.48}
\definecolor{mygray}{gray}{0.95}

\graphicspath{{}}

\begin{document}

%\title{Predictions of $m_{ee}$ for the neutrinoless double beta decay from a consistent Froggatt-Nielsen model}
\title{Predictions of $m_{ee}$ and neutrino mass from a consistent Froggatt-Nielsen model}

\author{Yu-Cheng Qiu}
\email{ethanqiu@sjtu.edu.cn}
\affiliation{Tsung-Dao Lee Institute and School of Physics and Astronomy, \\
Shanghai Jiao Tong University, 520 Shengrong Road, Shanghai, 201210, China}

\author{Jin-Wei Wang}
\email{jinwei.wang@uestc.edu.cn}
\affiliation{School of Physics, University of Electronic Science and Technology of China, Chengdu 611731, China}

\author{Tsutomu T. Yanagida}
\email{tsutomu.tyanagida@sjtu.edu.cn}
\affiliation{Tsung-Dao Lee Institute and School of Physics and Astronomy, \\
Shanghai Jiao Tong University, 520 Shengrong Road, Shanghai, 201210, China}
\affiliation{Kavli IPMU (WPI), The University of Tokyo, Kashiwa, Chiba 277-8583, Japan}

\begin{abstract}
The seesaw mechanism is the most attractive mechanism to explain the small neutrino masses, which predicts the neutrinoless double beta decay ($0\nu\beta\beta$) of the nucleus. Thus the discovery of $0\nu\beta\beta$ is extremely important for future particle physics. However, the present data on the neutrino oscillation is not sufficient to predict the value of $m_{ee}$ as well as the neutrino mass $m_\nu^i$. 
In this short article, 
by adopting a simple and consistent 
Froggatt-Nielsen model, which can well explain the observed masses and mixing angles of quark and lepton sectors, 
we calculate the distribution of $m_{ee}$ and $m_\nu^i$. Interestingly, a relatively large part of the preferred parameter space can be detected in the near future. 

\end{abstract}

\maketitle

\section{Introduction}

The Standard Model of particle physics contains $28$ free parameters, including the neutrino sector, which could not be explained theoretically. Among them, Yukawa couplings have a hierarchy structure, that is the flavor puzzle, which has attracted theorists' attention for decades \cite{Altarelli:2010gt,Ishimori:2010au,Hernandez:2012ra,King:2013eh,King:2014nza,King:2017guk,Feruglio:2019ybq,Altmannshofer:2022aml}. Meanwhile, for the neutrino sector, the neutrino oscillation experiments, e.g. Super-K \cite{Super-Kamiokande:1998kpq}, SNO \cite{SNO:2002tuh}, and Daya Bay \cite{DayaBay:2012fng}, have shown that neutrinos are not massless but possess tiny mass $m_\nu^i~(i = 1,2,3)$. Among different explanations, the seesaw mechanism~\cite{Minkowski:1977sc,Yanagida:1979as,Yanagida:1979gs,Gell-Mann:1979vob} is regarded as the most natural and promising one. An important corollary to the seesaw mechanism is the neutrinoless double beta decay ($0\nu\beta\beta$), which is closely related to the effective Majorana mass $m_{ee}$. Therefore, the discovery of $0\nu\beta\beta$ will be a huge breakthrough for the particle physics community. Nevertheless, neither the seesaw mechanism nor neutrino oscillation experiments can tell the values of $m_{ee}$ and $m_\nu^i$. This could be viewed as another intriguing puzzle. In this work, we try to propose a simple and consistent model that can well explain the fermion mass hierarchy and predict the values of $m_{ee}$ and $m_\nu^i$ simultaneously.

It is well known that the Forggatt-Nielsen (FN) mechanism~\cite{Froggatt:1978nt,Leurer:1993gy,Leurer:1992wg} provides an excellent method to explain the flavor puzzle, in which the Standard Model (SM) gauge group is extended by a horizontal global $U(1)_{\rm FN}$ symmetry. The $U(1)_{\rm FN}$ is broken by a vacuum expectation value (VEV) of a new scalar field $\phi$ whose $U(1)_{\rm FN}$ charge is $-1$. Naturally, a dimensionless parameter $\lambda$ can be defined, i.e. $\lambda= \langle \phi \rangle /M_{\rm PL} \sim \mathcal{O}(0.1)$, where $M_{\rm PL} \simeq 2.4\times 10^{18}\,{\rm GeV}$ is the reduced Planck scale. 
All SM particles also carry the $U(1)_{\rm FN}$ charge, and the value of the FN charge is generation dependent, which indicates that the masses of different generations of particles get suppressed by different powers of $\lambda$. 
Thus, the hierarchy issue could be well explained by the FN mechanism (see Ref. \cite{Altmannshofer:2022aml} for a very recent review).

Obviously, the core of the FN mechanism is the assignment of FN charge for SM particles. Recently, 
there are some works that did a blanket search to find optimal FN charge assignment \cite{Nishimura:2023wdu,Cornella:2023zme}. Especially, in Ref. \cite{Nishimura:2023wdu} the advanced reinforcement learning technics are involved. In our work, instead of adopting these kinds of brute force methods, we attempt to fix the FN charge of SM particles by doing a qualitative analysis. The rationality of the FN charge assignment is evaluated by comparing the theory predictions with experimental observations. In fact, it turns out that our strategy is quite effective, and to some extent, our results are in good agreement with previous blanket scan results~\cite{Cornella:2023zme}.

As for the more interesting neutrino sector, the seesaw mechanism actually implies that the neutrino masses can be produced from a
dimension-five effective operator \cite{Yanagida:1979gs,Weinberg:1979sa,Yanagida:1981xy}, which can be derived by integrating out the heavy right-handed neutrino states. In this case, the neutrino mass and mixing angle are also affected by the FN charge, since $\nu_L^i$ belongs to the electroweak doublet $\ell_{\rm L}^i$ and also carries FN charges.  Therefore, we show that it is possible to handle the flavor and neutrino puzzle within a unified FN framework. Interestingly, combined with the measurements of neutrino mass square difference $\Delta m_{21}^2$, $|\Delta m_{32}^2|$ \cite{Workman:2022ynf}, and some cosmological constraints on $\sum m_\nu^i$ \cite{eBOSS:2020yzd}, we can calculate the distribution of $m_{ee}$ and $m_\nu^i$. Surprisingly, we find that our predictions on $m_\nu^i$ are quite consistent with the available experimental data. Besides, a relatively large parameter space of $m_{ee}$ of our model could be explored in the near future neutrinoless $\beta\beta$ decay experiment, e.g. LEGEND-1000 \cite{LEGEND:2021bnm}. 

This paper is organized as follows. In Sec.~\ref{sec:fermion} we give a brief introduction to the FN mechanism and the analysis of how to fix the FN charges. In Sec.~\ref{sec:neutrino} we calculate the predictions of our model on $m_{ee}$ and $m_\nu^i$ as well as the near future constraints from LEGEND-1000. Conclusions and further discussions are given in Sec.~\ref{sec:summary}.

%the key We adopt a scheme called as ``minimal FN model"~\cite{Fedele:2020fvh} in which all fermion mass matrices are determined basically by FN charges of the fermions and the parameter $\lambda$. Here, the neutrino mass matrix is given by the dimension 5 operator~\cite{Yanagida:1979gs,Weinberg:1979sa,Yanagida:1981xy}.

%We have searched for the best set of the FN charges of quarks and lepton, given in Table~\ref{tab:FN_charges}.
%The strategy for determining the FN charges is the following.
%~\footnote{One could also perform a blanket search of FN charges as in Ref.~\cite{Nishimura:2023wdu,Cornella:2023zme}.} 
%As a starting point, 
%since one could always rescale $\lambda$ by adding an overall charge to all generations, 
%we first fix the FN charge of the top quark, $\overline{Q}_{\rm L}^3$, to be zero.

%Some details of charge determination are explained in the appendix~\ref{appendix:FN_charges}.

\section{The consistent Forggatt-Nielsen model}\label{sec:fermion}

The FN model we are considering is a simple extension of the Standard Model~\cite{Froggatt:1978nt,Leurer:1993gy,Leurer:1992wg}. The mass matrices of quarks are granted by Yukawa couplings, which are
\begin{equation}
    -\mathcal{L}\supset  y_u^{ij} \overline{Q}_{\rm L}^i \tilde{H} u_{\rm R}^j +  y_d^{ij}\overline{Q}_{\rm L}^i H d_{\rm R}^j + {\rm  h.c.}\;,
    \label{eq:quarkLag}
\end{equation}
where $H$ is the Standard Model Higgs doublet and $\tilde{H}=i\sigma_2 H^*$.
Under our FN framework, the Yukawa couplings can be expressed as
\begin{equation}
    y_u^{ij} = g_{ij} \lambda^{n_u^{ij}} \;,\quad y_d^{ij}=g_{ij} \mathcal{N} \lambda^{n_d^{ij}}\;,
    \label{eq:yd}
\end{equation}
where the $\mathcal{N}$ is an overall factor to accommodate the overall scale difference between up-type and down-type sectors, whose origin could be two-Higgs-doublet models at high energy~\cite{Branco:2011iw}, the $g_{ij}$ is the universal coupling, whose magnitude $|g_{ij}|$ fulfills a normal distribution $N(\mu,\sigma^2)$, while its argument fulfills a uniform distribution from $0$ to $2\pi$. In our work, we choose $\mu=1$ and $\sigma=0.3$ for a benchmark case. Since we have set $U(1)_{\rm FN}^H=0$ and $U(1)_{\rm FN}^\phi=-1$, the value of $n_{u/d}^{ij}$ is determined by the FN charge of quarks,
\begin{subequations}
\begin{align}
    n_{u}^{ij} &= U(1)_{\rm FN}^{\overline{Q}_{\rm L}^i} + U(1)_{\rm FN}^{u_{\rm R}^j}\;,\\
    n_{d}^{ij} &= U(1)_{\rm FN}^{\overline{Q}_{\rm L}^i} + U(1)_{\rm FN}^{d_{\rm R}^j}\;.
\end{align}
\end{subequations}
Clearly, once we fix the value of $\lambda$ and quarks' FN charge, the quark mass, mixing angle, and $CP$ angle are almost fixed. 
Mass hierarchy is indicated by the fermion mass ratio between generations,
i.e. $m_u/m_t$, $m_d/m_b$, and so on. We find that if we focus on the mass ratios, then the charge assignment will become much easier.
Take $\overline{Q}_{\rm L}^i$ and $d_{\rm R}^j$ for example, where the most general form of their FN charge should be $U(1)_{\rm FN}^{\overline{Q}_{\rm L}^i}=\{a,b,c\}$ and $U(1)_{\rm FN}^{d_{\rm R}^j}=\{d,e,f\}$, then we have
\begin{equation}
    \lambda^{n_d^{ij}} = \lambda^{c+f}\begin{pmatrix}
        \lambda^{a-c+d-f} & \lambda^{a-c+e-f} & \lambda^{a-c} \\
        \lambda^{b-c+d-f} & \lambda^{b-c+e-f} & \lambda^{b-c} \\
        \lambda^{d-f} & \lambda^{e-f} & 1 \\
    \end{pmatrix}\;.
    \label{eq:lambdafactor}
\end{equation}
The overall factor $\lambda^{c+f}$ will not affect the fermion mass ratio and mixings. However, this factor could also be used to explain the absolute quark mass (such as $\mathcal{N}$ in Eq.~\eqref{eq:quarkLag}), i.e. the mass hierarchy between up-type and down-type quarks. In the following content, we just set the FN charge of third-generation fermion equal to zero, which is equivalent to absorbing the $\lambda^{c+f}$ factor into $\mathcal{N}$, and we will comment on this issue in Sec.~\ref{sec:summary}. For simplicity, we only consider the FN charge to be an integer or half-integer less than 5. Similar conditions are also adopted in previous literature~\cite{Nishimura:2023wdu,Cornella:2023zme}.

For the quark sector, one could show that the Cabibbo-Kobayashi-Maskawa (CKM) matrix is mainly determined by the $\overline{Q}_{\rm L}^i$. 
The well-known Wolfenstein parametrization~\cite{Wolfenstein:1983yz} indicates that mixing angles in the CKM matrix approximately satisfy $\sin\theta_{12}^{\rm C} \sim \lambda'$, $\sin\theta_{23}^{\rm C}\sim \lambda'^2$ and $\sin\theta_{13}^{\rm C} \sim \lambda'^3$, where $\lambda'\sim 0.2$. It is attempted to assume that $\lambda \sim \mathcal{O}(\lambda')$, and the FN charge of the quark doublet shall be $U(1)_{\rm FN}^{\overline{Q}_{\rm L}^i}=\{3,2,0\}$ to produce such a mixing pattern~\cite{Fedele:2020fvh, Cornella:2023zme}. 
Once we know the $U(1)_{\rm FN}^{\overline{Q}_{\rm L}^i}$, the FN charge of $u_{\rm R}$ and $d_{\rm R}$ can be roughly fixed by comparing with the observed quark mass ratios.

In Table~\ref{tab:experiments} we have summarized all available mass ratios, mixing angles, and $CP$ angles of quark and lepton sectors. Considering that the $U(1)_{\rm FN}$ was broken at a very high energy scale $\sim M_{\rm PL}$, all the numbers in Table~\ref{tab:experiments} should also be evaluated at a high energy scale. From Refs.~\cite{Huang:2020hdv,Martin:2019lqd}, we can see that the mass ratio of the quark and lepton sectors are almost energy independent as long as the energy scale larger than $\sim 10^8\,{\rm GeV}$. Therefore, we can safely substitute the mass ratio at $\sim 10^{12}\, {\rm GeV}$ for the results at $\sim M_{\rm PL}$. As for the mixing angles and $CP$ angle, we assume that they are energy independent.
\begin{table}
    \caption{
    Experimental measured quantities. Quark and lepton mass ratios are taken at the scale of $10^{12}\,{\rm GeV}$~\cite{Huang:2020hdv}, which are almost energy scale independent~\cite{Martin:2019lqd}. Mixing angles and CP phases are in the rad unit~\cite{ParticleDataGroup:2022pth}.}
    \begin{ruledtabular}
        \begin{tabular}{c c c c}
            $m_u/m_t$ & $m_c/m_t$ & $m_d/m_b$ & $m_s/m_b$ \\
            $6.58\times 10^{-6}$ & $0.00333$ & $0.00104$ & $0.0201$\\
            \midrule
            $\theta_{12}^{\rm C}$ & $\theta_{23}^{\rm C}$ & $\theta_{13}^{\rm C}$ & $\delta^{\rm C}$ \\
              $0.227$ & $0.0418$ & $0.00369$ &  $1.14$        \\
             \midrule
             $m_e/m_\tau$ & $m_\mu/m_\tau$ & $\Delta m_{21}^2/\Delta m_{32}^2$ & \\
             $0.000279$ & $0.0589$ & $0.0307$ & \\
             \midrule
              $\theta_{12}^{\rm P}$ & $\theta_{23}^{\rm P}$ & $\theta_{13}^{\rm P}$ & $\delta^{\rm P}$  \\
             $0.591$ & $0.844$ & $0.150$ & $-2.41_{-0.489}^{+0.663}$
        \end{tabular}
    \end{ruledtabular}
    \label{tab:experiments}
\end{table}

Assuming the FN charges of $u_{\rm R}$ and $d_{\rm R}$ are $\{a,b,0\}$ and $\{c,d,0\}$, respectively, we can derive that
\begin{equation}
    n_u= 
    \begin{pmatrix}
        3+a & 3+b & 3 \\
        2+a & 2+b & 2 \\
        a & b & 0 \\
    \end{pmatrix}\;,
    n_d= 
    \begin{pmatrix}
        3+c & 3+d & 3 \\
        2+c & 2+d & 2 \\
        c & d & 0 \\
    \end{pmatrix}\;.
    \label{eq:udquark}
\end{equation}
From Table~\ref{tab:experiments} we roughly have 
\begin{equation}
    \frac{m_u}{m_t} \sim \lambda'^7 \;,~ \frac{m_c}{m_t} \sim \lambda'^{3.5} \;,~ \frac{m_d}{m_b} \sim \lambda'^4 \;,~ \frac{m_s}{m_b} \sim \lambda'^2\;,
    \label{eq:quarkratio}
\end{equation}
where the $\lambda' \sim 0.2$. Comparing Eq.~\eqref{eq:udquark} and Eq.~\eqref{eq:quarkratio}, we can derive that 
\begin{equation}
    a=4\;, \quad b=1.5 \;, \quad c=1\;, \quad d=0\;.
\end{equation}

For the lepton sector, the mass terms are generated by Yukawa couplings and a five-dimensional operator, that is 
\begin{equation}
    -\mathcal{L}\supset y_\ell^{ij} \overline{\ell}_{\rm L}^i H e_{\rm R}^j + \frac{1}{M} y_\nu^{ij} \left(\overline{\ell^{\rm c}_{\rm L}}^i \tilde{H}^*\right) \left( \tilde{H}^\dagger \ell^j_{\rm L} \right) + {\rm h.c.}\;,
    \label{eq:lepton_yukawa}
\end{equation}
where
\begin{equation}
    y_\ell^{ij} =  g_{ij} \mathcal{N}\lambda^{n_\ell^{ij}} \;, \quad y_\nu^{ij} = g_{ij}' \lambda^{n_\nu^{ij}}\;.
    \label{eq:quark_yukawa}
\end{equation}
Note that here we use the same $\mathcal{N}$ as Eq.~\eqref{eq:yd} because of the fact that $m_b\sim m_\tau$ at a very high energy scale \cite{Huang:2020hdv}. Besides, the $g_{ij}'$ is a symmetric matrix due to the Majorana nature of the neutrino. Similar to the quark sector, the Pontecorvo-Maki-Nakagawa-Sakata (PMNS) matrix is mainly determined by $\overline{\ell}_{\rm L}^i$. The observations tell us that 
the mixing angles $\theta_{12}^{\rm P}$ and $\theta_{23}^{\rm P}$ are relatively larger compared to $\theta_{13}^{\rm P}$, and $\theta_{12}^{\rm P}$ is slightly smaller than $\theta_{23}^{\rm P}$. 
Following a similar logic and using the universal $\lambda'$, we assign the FN charge for the lepton doublet as $U(1)_{\rm FN}^{\overline{\ell}_{\rm L}}=\{1,0.5,0\}$. 
One interesting fact is that we can easily prove that the rank of $n_\nu$ is 1, which means
there would be two mass eigenvalues being almost zero and one relatively large eigenvalue after diagonalization. This indicates that
the FN mechanism naturally prefers normal order (NO)\footnote{Another perspective to understand this feature is through the FN charge assignment of $\overline{\ell}_{\rm L}$. Combining Eq. (8) and $U(1)_{\rm FN}^{\overline{\ell}_{\rm L}}=\{1,0.5,0\}$  (see Table \ref{tab:FN_charges}), the ratio of three eigenvalues of $y_\nu^{ij}$ is roughly $1:\lambda':\lambda'^2$, which is clearly NO for $\lambda'\sim 0.2$.}. Therefore, in the following content, we just stick to the NO scenario.

Then $U(1)_{\rm FN}^{e_{\rm R}}$ is obtained from estimating the charged lepton mass ratios. Specifically, assuming $U(1)_{\rm FN}^{e_{\rm R}}=\{e,f,0\}$ we have
\begin{equation}
    n_\ell= 
    \begin{pmatrix}
        1+e & 1+f & 1 \\
        0.5+e & 0.5+f & 0.5 \\
        e & f & 0 \\
    \end{pmatrix}\;.
    \label{eq:electron}
\end{equation}
From Table.~\ref{tab:experiments} we roughly have
\begin{equation}
    \frac{m_e}{m_\tau} \sim \lambda'^5 \;,\quad \frac{m_\mu}{m_\tau} \sim \lambda'^{1.5}\;,
    \label{eq:electronratio}
\end{equation}
where the $\lambda' \sim 0.2$. Comparing Eq.~\eqref{eq:electron} and Eq.~\eqref{eq:electronratio}, we can derive that 
\begin{equation}
    e=4\;,\quad f=1\;.
\end{equation}

% one may first assume that $\overline{\ell}_{\rm L}$ carry FN charge $(1,0,0)$. However, since $\theta_{12}^{\rm P}$ slightly smaller than $\theta_{23}^{\rm P}$, we assign the FN charge for the lepton doublet as $\overline{\ell}_{\rm L}(1,0.5,0)$. 

\begin{table}
    \caption{FN charge of quarks and leptons.}
    \begin{ruledtabular}
    \begin{tabular}{c c c c}
        Generation $i$ &  $1$ & $2$ & $3$\\
        \midrule
       $\overline{Q}_{\rm L}$  &  $3$  &  $2$  & $0$ \\
       $u_{\rm R}$  & $4$ & $1.5$ & $0$ \\
       $d_{\rm R}$  & $1$ &  $0$ & $0$ \\
       $\overline{\ell}_{\rm L}$ & $1$ & $0.5$  & $0$ \\
       $e_{\rm R}$  &  $4$ & $1$ & $0$ \\
    \end{tabular}
    \end{ruledtabular}
    \label{tab:FN_charges}
\end{table}

Until now, by doing a qualitative analysis we have fixed the FN charge of SM particles (see Table~\ref{tab:FN_charges}). 
However, there are two issues that need to be emphasized. The first one is the global $U(1)_{\rm FN}$ symmetry. The above analysis is based on an assumption that all terms in Eq.~\eqref{eq:quarkLag} and Eq.~\eqref{eq:lepton_yukawa} respect the global $U(1)_{\rm FN}$. However, it is believed that any
global symmetries must be broken by nonperturbation effects in quantum gravity~\cite{Banks:2010zn}. With the charge assignment in Table~\ref{tab:FN_charges}, we found there is a discrete $\mathbf{Z}_{33}$ symmetry, actually this symmetry is anomaly-free for $\mathbf{Z}_{33}\times [SU(2)_{\rm L}]^2$ and $\mathbf{Z}_{33}\times [SU(3)_{\rm c}]^2$, and therefore can be gauged. This can be regarded as quite an interesting feature of our model.

%The global $U(1)_{\rm FN}$ symmetry could easily be explicitly broken by at least quantum gravity. Here we use gauged discrete subgroup of the FN $U(1)$ symmetry, $\mathbf{Z}_{33}$, with $0.5$ as the charge unit in Table~\ref{tab:FN_charges}. One could see that such charge assignment is anomaly-free, concerning $\mathbf{Z}_{33}\times [SU(2)_{\rm L}]^2$ and $\mathbf{Z}_{33}\times [SU(3)_{\rm c}]^2$.

Another issue is the exact value of $\lambda$, which is the only free parameter after fixing the FN charge. We conduct the analysis by using a very rough number, i.e. $\lambda \sim \lambda'\sim0.2$, while a more accurate $\lambda$ is necessary for a concrete FN model. In the following content, we adopt the minimum chi-square method to find the best value of $\lambda$.

The strategy is quite straightforward. As we mentioned above, in our model the $g_{ij}$ is the universal coupling whose magnitude fulfills a normal distribution $N(1,0.3)$, while its argument fulfills a uniform distribution from 0 to $2\pi$. With the fixed FN charges and $\lambda$, all the couplings, e.g., $y_u^{ij}$, $y_d^{ij}$, $y_\ell^{ij}$, $y_\nu^{ij}$, can be generated. Then, all the desired quantities (denoted $X_i$), including the quark and lepton mass ratios, the mixing angles, and $CP$ angles, will be fixed. By randomly generating $g_{ij}$ we can derive the distribution $X_i$. The chi square is defined as
\begin{equation}
    \chi^2(\lambda) = \sum_i\left( \frac{E(X_i)-X_i^{\rm exp}}{\sqrt{V(X_i)}}\right)^2\;,
    \label{eq:chi2}
\end{equation}
where $X_i^{\rm exp}$ is the experimentally measured value (see Table~\ref{tab:experiments}), $E(X_i)$ is the expectation value of $X_i$, while $V(X_i)$ is the deviation. Here we take
\begin{align*}
    X_i&\in \Big\{ \frac{m_u}{m_t},\frac{m_c}{m_t},\frac{m_d}{m_b},\frac{m_s}{m_b},\frac{m_e}{m_\tau},\frac{m_\mu}{m_\tau},\frac{\Delta m_{21}^2}{\Delta m_{32}^2},\\
    &\qquad \qquad  \theta_{12}^{\rm C},\theta_{23}^{\rm C},\theta_{13}^{\rm C},\delta^{\rm C},\theta_{12}^{\rm P},\theta_{23}^{\rm P},\theta_{13}^{\rm P}\Big\}\;,
\end{align*}
where $\delta^{\rm P}$ is not the direct observable. By scanning the parameter space of $\lambda$ we can find the best value that minimizes the $\chi^2(\lambda)$. 

\begin{figure*}
\centering
\includegraphics[width=0.99\textwidth]{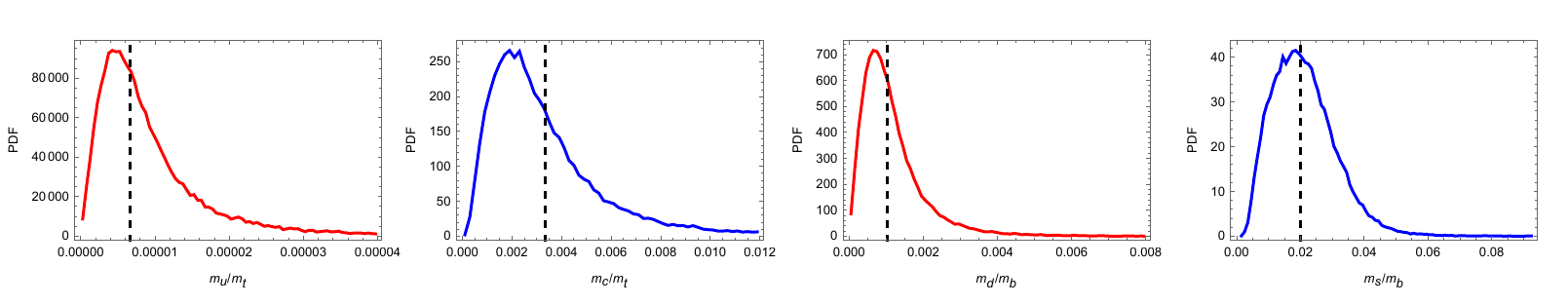}
\caption{The PDF of quark mass ratios with $\lambda=0.171$ and $\sigma=0.3$. The black dashed vertical lines indicate the experimental measurements given in Table~\ref{tab:experiments}.}
\label{fig:quark_mass}
\end{figure*}
\begin{figure*}
\centering
\includegraphics[width=0.99\textwidth]{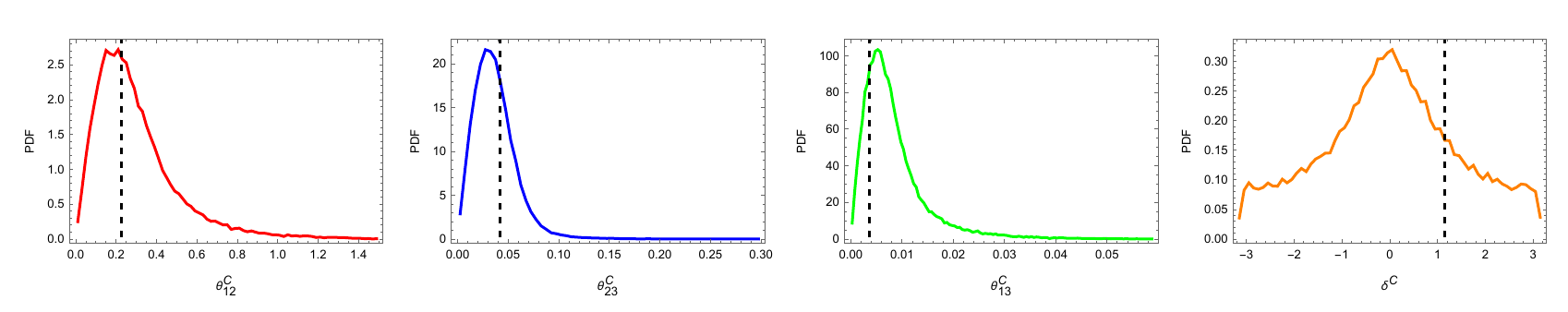}
\caption{The PDFs of the mixing angle $\theta_{ij}^\text{C}$ and $CP$ angle $\delta^\text{C}$ of quark sector with $\lambda=0.171$ and $\sigma=0.3$. The black dashed vertical lines indicate the experimentally measured value as indicated in Table~\ref{tab:experiments}.}
\label{fig:ckm}
\end{figure*}

To calculate Eq.~\eqref{eq:chi2} we need the exact distribution of $X_i$. For the quark sector, the Yukawa matrices can be decomposed as
\begin{equation}
    y_u = U_u D_u W_u^\dagger\;,\quad y_d = U_d D_d W_d^\dagger\;,
    \label{eq:quark_decomposition}
\end{equation}
where $U_{u,d}$ and $W_{u,d}$ are unitary matrices, $D_{u,d}$ is a diagonal matrix with all real elements. The $U_u$ and $W_u$ are obtained from
$y_u y_u^\dagger = U_u (D_u)^2 U_u^\dagger$ and $y_u^\dagger y_u = W_u (D_u)^2 W_u^\dagger$, and the same is true for the $U_d$ and $W_d$. Then up-type and down-type quark mass ratios are
\begin{align}
    \frac{m_u}{m_t}&=\frac{D_u^{11}}{D_u^{33}}\;,\quad \frac{m_c}{m_t}=\frac{D_u^{22}}{D_u^{33}}\;, \nonumber \\
    \frac{m_d}{m_b}&=\frac{D_d^{11}}{D_d^{33}}\;,\quad \frac{m_s}{m_b}=\frac{D_d^{22}}{D_d^{33}}\;.
    \label{eq:quarkmass_ratio}
\end{align}
Within our notation, the CKM matrix can be expressed as
\begin{equation}
    U_{\rm CKM} = U_u^\dagger U_d\;,
    \label{eq:ckm}
\end{equation}
which contains all information on quark mixing angles ($\theta_{ij}^\text{C}$) and $CP$ angle ($\delta^\text{C}$).
Under standard parametrization, the $U_{\rm CKM}$ is 
\begin{equation}
    U_{\rm CKM} = V_{\rm SP}(\theta_{12}^{\rm C},\theta_{23}^{\rm C},\theta_{13}^{\rm C},\delta^{\rm C})\;,
    \label{eq:ckm_decomposition}
\end{equation}
where $V_{\rm SP}$ is a unitary matrix possessing four real parameters, that is
\begin{widetext}
    \begin{equation}
         V_{\rm SP}(\theta_{12},\theta_{23},\theta_{13},\delta) =
            \begin{pmatrix}
                c_{13}c_{12} & c_{13}s_{12} & s_{13} e^{-i\delta}\\
                -c_{23}s_{12}-s_{23}s_{13}c_{12} e^{i\delta} & c_{23} c_{12}-s_{23}s_{13}s_{12} e^{i\delta} & s_{23}c_{13} \\
                s_{23}s_{12}-c_{23}s_{13}c_{12} e^{i\delta} & -s_{23}c_{12}-c_{23}s_{13}s_{12} e^{i\delta} & c_{23} c_{13} 
            \end{pmatrix}\;, 
            \label{eq:CKM}
    \end{equation}
\end{widetext}
where $c_{ij} = \cos\theta_{ij}$ and $s_{ij} = \sin \theta_{ij}$. Note that we adopt the convention that $\theta_{ij}\in[0,\pi/2)$ and $\delta\in[-\pi,\pi)$. 
Utilizing Eq.~\eqref{eq:quarkmass_ratio} and Eq.~\eqref{eq:CKM}, we can calculate quark sector parameters.

\begin{figure*}
\centering
\includegraphics[width=0.8\textwidth]{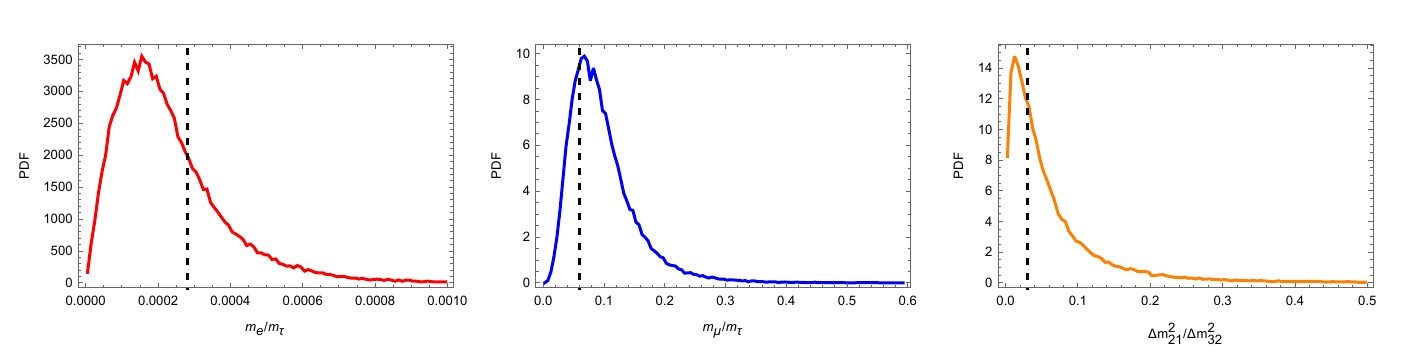}
\caption{Similar to Fig.~\ref{fig:quark_mass} but the lepton sector. Note that for neutrinos the ratio of mass square difference, i.e., $\Delta m_{21}^2/\Delta m_{32}^2$ is shown.}
\label{fig:lepton_mass}
\end{figure*}
\begin{figure*}
\centering
\includegraphics[width=0.99\textwidth]{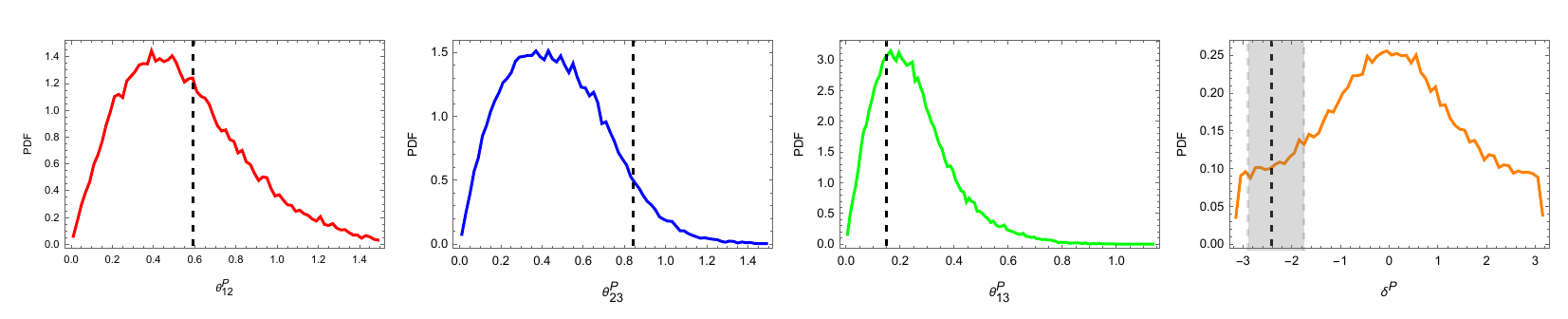}
\caption{Similar to Fig.~\ref{fig:ckm} but for the neutrino sector. The Gray dashed vertical lines indicate the experimentally measured values as indicated in Table~\ref{tab:experiments}.}
\label{fig:pmns}
\end{figure*}

One significant difference for the lepton sector (see Eq.~\eqref{eq:lepton_yukawa}) is that the neutrino mass is generated by a five-dimensional effective operator. As we mentioned in this case $y_\nu$ is a complex symmetric matrix. 
Similar to the quark sector, we do the following decomposition, i.e.,
\begin{equation}
    y_\ell = U_\ell D_\ell W_\ell^\dagger\;,\quad y_\nu = U_\nu  D_\nu U_\nu^T\;,
    \label{eq:lepton_decomposition}
\end{equation}
where $D_\ell$ and $D_\nu$ are diagonal matrices with all real elements and $U_\ell$, $W_\ell$, and $U_\nu$ are unitary matrices. The $U_\ell$ and $W_\ell$ are obtained from $y_\ell y_\ell^\dagger = U_\ell (D_\ell)^2 U_\ell^\dagger$ and $y_\ell^\dagger y_\ell = W_\ell (D_\ell)^2 W_\ell^\dagger$. 
Then the mass ratios for charged leptons are
\begin{equation}
    \frac{m_e}{m_\tau}=\frac{D_\ell^{11}}{D_\ell^{33}}\;, \quad \frac{m_\mu}{m_\tau}=\frac{D_\ell^{22}}{D_\ell^{33}}\;.
\end{equation}
Next, we need to derive the explicit form of $U_\nu$. Define $\tilde{U}_\nu$ such that $y_\nu^\dagger y_\nu = \tilde{U}_\nu (D_\nu)^2 \tilde{U}_\nu^\dagger$, and since $y_\nu$ is a symmetric matrix, we can derive that $y_\nu=\tilde{U}_\nu \Phi D_\nu \tilde{U}_\nu^T$, where $\Phi$ is a diagonal matrix and each element is a pure phase. Compared with Eq.~\eqref{eq:lepton_decomposition} we have $U_\nu = \tilde{U}_\nu \Phi^{-1/2}$.
In our notation, the PMNS matrix can be written as 
\begin{equation}
    U_{\rm PMNS} = U_\ell^\dagger U_\nu^* \;,
    \label{eq:pmns}
\end{equation}
Different from the CKM matrix, one could only rotate three phases from charged leptons, which results in two extra phases in $U_{\rm PMNS}$ compared to the CKM matrix,
\begin{equation}
    U_{\rm PMNS} = V_{\rm SP} (\theta_{12}^{\rm P}, \theta_{23}^{\rm P}, \theta_{13}^{\rm P}, \delta^{\rm P}) 
    \begin{pmatrix}
        1 & 0 & 0 \\
        0 & e^{i \frac{\alpha_{\rm M}}{2}} & 0 \\
        0 & 0 & e^{i\frac{\beta_{\rm M}}{2}} 
    \end{pmatrix}\;,
    \label{eq:pmns_decomposition}
\end{equation}
where $\alpha_{\rm M}$ and $\beta_{\rm M}$ are the Majorana phases. In our work, we only consider $\theta_{ij}^\text{P}$ and $\delta^\text{P}$.

So now we know how to calculate $X_i$, and then by repeating random sampling $g_{ij}$ and $g_{ij}'$, we can get the distribution of $X_i$.
Combining with Eq.~\eqref{eq:chi2} we do a parameter scan to get the best value of $\lambda$, that is
\begin{equation}
    \lambda=0.171\; ~~~\text{with}~~~ \chi^2=4.69\;.
    \label{eq:lambda}
\end{equation}

Under such an input, the probability density function (PDF) of the quark mass ratios are plotted in Fig.~\ref{fig:quark_mass}, and the results of mixing angles and $CP$ angle are shown in Fig.~\ref{fig:ckm}. The black dashed lines indicate the experimental measurements (see Table~\ref{tab:experiments}). It shows that our model predictions all agree with the experimental observations. 
Similarly, Fig.~\ref{fig:lepton_mass} and Fig.~\ref{fig:pmns} show the PDF of the lepton mass ratio and mixing angles. For neutrinos, there are only mass square differences available, i.e. $\Delta m_{21}^2=(m_\nu^2)^2-(m_\nu^1)^2$ and $\Delta m_{32}^2=(m_\nu^3)^2-(m_\nu^2)^2$. Note that $\delta^\text{P}$ is not a directly observable quantity and has a large uncertainty, so it is not included in $\chi^2(\lambda)$.
In conclusion, adopting the FN charge in Table~\ref{tab:FN_charges} and $\lambda=0.171$, our FN model could successfully explain $15$ parameters in the standard model.
In comparison, we also calculate the $\chi^2$ by using the best three charge assignments in Table. I of Ref.~\cite{Cornella:2023zme}, and we find that our model has a smaller $\chi^2$.
\begin{figure}
    \centering
\includegraphics[width=0.35\textwidth]{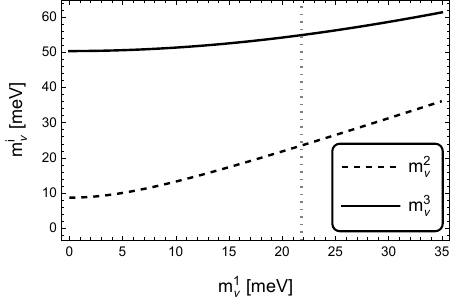}
    \caption{Neutrino mass spectrum under the constraint from cosmology and oscillation measurements. The gray vertical dotted line indicates the upper bound on $m_1$, above which the cosmological bound, i.e. Eq.~\eqref{eq:mi_sum}, would be violated.}
    \label{fig:m3}
\end{figure}
\begin{figure*}
\centering
\includegraphics[width=0.8\textwidth]{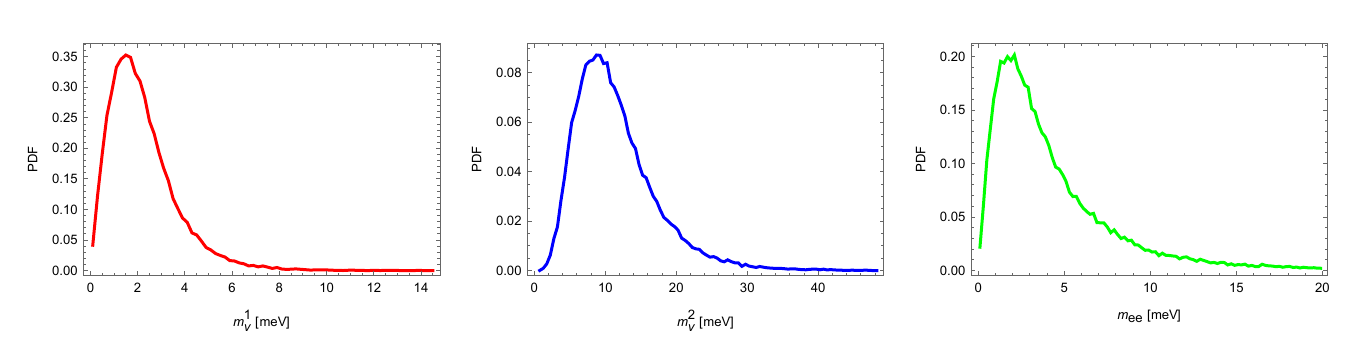}
\caption{The PDF of neutrino mass $m^{1,2}_\nu$ and effective Majorana mass element $m_{ee}$. Note that $m_3=0.05\,{\rm eV}$ is adopted due to the cosmology and oscillation experiments constraints (see Fig.~\ref{fig:m3}).}
\label{fig:neutrino_mass}
\end{figure*}
\begin{figure}
\centering
\includegraphics[width=0.35\textwidth]{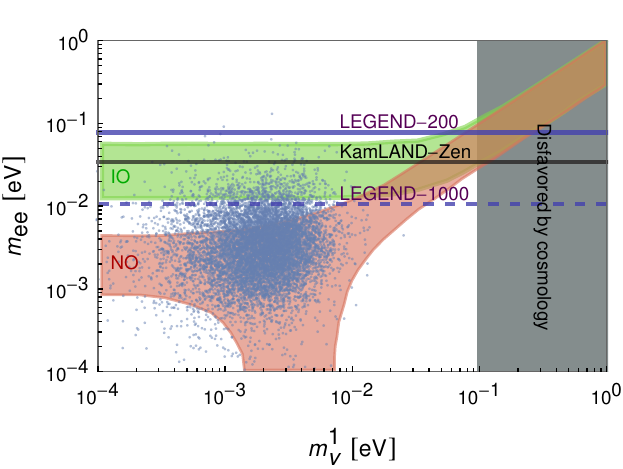}
\caption{The scattering plot (blue points) on  $m_\nu^1$--$m_{ee}$ plane with $\lambda=0.171$ and $\sigma=0.3$. The horizontal lines indicate different experiment constraints, e.g. KamLAND-Zen~\cite{KamLAND-Zen:2022tow}, LEGEND-200~\cite{GERDA:2020xhi}, and LEGEND-1000~\cite{LEGEND:2021bnm}.}
\label{fig:mee_m1}
\end{figure}

\section{Predictions on $m_\nu^i$ and $m_{ee}$}\label{sec:neutrino}

In Sec.~\ref{sec:fermion}, we have built a consistent FN model, and its predictions of 15 parameters agree very well with experimental observations. Based on this success, we are more interested in its predictions on the neutrino sector, especially, $m_\nu^i$ and $m_{ee}$.
As we mentioned in Sec.~\ref{sec:fermion},
the FN mechanism naturally prefers the NO.
As a cross-check, we randomly generate a neutrino Yukawa matrix, $y_\nu$, whose eigenvalues are $\{D_\nu^i\}$. If the average of $\{D_\nu^1, D_\nu^2, D_\nu^3 \}$ is smaller than the median, $y_\nu$ is inverted order (IO). By sampling $10^6$ times using our FN charge with $\lambda=0.171$ and $\sigma=0.3$, we find that NO takes up $\sim 98\%$. Therefore, we only consider NO in this work.

The neutrino oscillation experiments can tell the neutrino mass difference, e.g. $\Delta m_{21}^2$ and $\Delta m_{32}^2$ (see Table.~\ref{tab:experiments}). Besides, the cosmology observations put constraints on total neutrino mass~\cite{ParticleDataGroup:2022pth}
\begin{equation}
    \sum_i m_\nu^i < 0.1\,{\rm eV}\;.
    \label{eq:mi_sum}
\end{equation}
Under these constraints, we can derive the neutrino mass spectrum (see Fig.~\ref{fig:m3}). 
The x axis represents $m_\nu^1$, while the black dashed and solid lines represent $m_\nu^2$ and $m_\nu^3$ respectively. The vertical dotted line indicates the cosmological bound given by Eq.~\eqref{eq:mi_sum}. One could see that $m_\nu^3$ is almost a constant. 
Thus, we can take $m_\nu^3=0.05\,{\rm eV}$ as a benchmark value to fix the neutrino mass normalization of our model. Specifically, by diagonalizing $y_\nu$ our model can predict $m_\nu^1/m_\nu^3$ and $m_\nu^2/m_\nu^3$. After we fix $m_\nu^3$, we can get the PDF of $m_\nu^1$ and $m_\nu^2$, which are shown in Fig.~\ref{fig:neutrino_mass}. It shows the most probable value of $m_\nu^1$ and $m_\nu^2$ are $\sim1.6$ meV and $\sim 9$ meV respectively, which is perfectly consistent with Fig.~\ref{fig:m3}.

%FN charges and $\lambda$~in \eqref{eq:lambda} are fixed by the last section, which gives us the mass ratio, $m_\nu^1/m_\nu^3$ and $m_\nu^2/m_\nu^3$, from randomly generated $y_\nu$. This results in predictions on neutrino masses, which are shown in Fig.~\ref{fig:neutrino_mass}.

%If the neutrino is Majorana fermion, then the dim-5 operator is sufficient to describe it~\eqref{eq:lepton_yukawa}. 
%With the success of the last section, we use this model to probe quantities that have not been observed in the neutrino sector.
%Two Majorana phases could be explicitly extracted once we generate the $U_{\rm PMNS}$. However, due to our assumption on the flat distribution of ${\rm arg}(g_{ij})$, our model has no preference for two Majorana phases, both of which end up with a flat distribution.
%Note that here we only use the NO data for the neutrino sector in Table~\ref{tab:experiments} and in the last section. This is because our model largely favors NO. To see it, 

% \begin{figure}
%     \centering
%     \includegraphics[width=6.5cm]{fig_mee.pdf}
%     \caption{Distribution of predicted $m_{ee}$ under $\lambda=0.171$ and $\sigma=0.3$ with $m_3=0.05\,{\rm eV}$. The black solid line is a fitted distribution using log-scale Gaussian distribution.}
%     \label{fig:mee}
% \end{figure}

As we mentioned the seesaw mechanism implies that the $0\nu\beta\beta$ process could happen. The decay rate is proportional to $m_{ee}^2$, 
%which is defined under the frame that charged leptons have been diagonalized to their mass eigenstates, which is
% \begin{equation}
%     -\mathcal{L}\supset  \overline{e}_{\rm L'}^k m_\ell^k e_{\rm R}'^k + \nu'^i (U_{\rm PMNS}^*)^{ij} m_\nu^j (U_{\rm PMNS}^\dagger)^{jk}  \nu'^k \;,
% \end{equation}
% where $m_\ell^k$ and $m_\nu^j$, with only one index on top, are diagonal mass matrix elements.
where $m_{ee}$ measures the transition amplitude for an electron neutrino to an electron neutrino while violating the lepton number, which could be expressed as
\begin{equation}
    m_{ee} = \left| \sum_i (U_{\rm PMNS}^*)_{1i}^2 m_\nu^i \right| \;.
\end{equation}
In the rightmost plot of Fig.~\ref{fig:neutrino_mass} we demonstrate the PDF of $m_{ee}$. We find that most probable value of $m_{ee}$ is located at $\sim 2\times 10^{-3}\,{\rm eV}$, and $m_{ee}$ is bounded by
\begin{equation}
    0.503 \,{\rm meV} \lesssim m_{ee} \lesssim 18.0\,{\rm meV}\;,
\end{equation}
at $95\%$ C.L..  Besides, in Fig.~\ref{fig:mee_m1}, we show the model prediction (blue dots) on the $m_\nu^1$-$m_{ee}$ plane. The green and red regions represent the IO and NO cases, respectively. The gray region is excluded by cosmological observation~\cite{ParticleDataGroup:2022pth}. Clearly, it shows most of the predictions overlap with the NO region. The near future $0\nu \beta \beta$ experiment LEGEND-1000 can check the validity of our modem to some extent \cite{LEGEND:2021bnm}.

\section{Summary and Discussion}\label{sec:summary}

In this short article, we propose a consistent FN model to deal with flavor puzzles and neutrino puzzles at the same time. One intriguing feature of this work is that we derive the FN charge of SM particles (see Table.~\ref{tab:FN_charges}) by doing a qualitative analysis instead of a brutal search. Then, there is only one free parameter $\lambda$ in our FN model, whose best value, i.e. $\lambda=0.171$, is fixed through the minimum chi-square estimation. 
By randomly sampling $g_{ij}$, we calculate the PFDs of mass ratios, mixing angles, and $CP$ angles (see Fig.~\ref{fig:quark_mass} -- Fig.~\ref{fig:pmns}). All these predictions agree with experimental observations quite well. 
We find that our charge assignments share some similarities with previous literature, e.g. Ref.~\cite{Cornella:2023zme}, however, our FN model has a smaller $\chi^2(\lambda)$ and therefore fits the experimental observations better.

Another striking feature is that our model possesses a discrete $\mathbf{Z}_{33}$, which is anomaly-free and can be gauged. Unlike the global $U(1)_\text{FN}$ symmetry, the gauged $\mathbf{Z}_{33}$ symmetry is free of quantum gravity corrections. Therefore, one could replace the $U(1)_\text{FN}$ with this gauged $\mathbf{Z}_{33}$, which makes our model more robust but would not change the existing conclusions.

%search which uses an anomaly-free discrete gauge $\mathbf{Z}_{33}$ symmetry, to ensure the integrity of the hierarchy structure. Remarkably, after the FN charges are fixed, this model has only one free parameter, $\lambda$, and it could explain $15$ experimental measurements.
%This model also provides us a prediction on $m_{ee}$ that is essential in the $0\nu\beta\beta$ process. We predict that $m_{ee}$ has a statistical upper and lower bound even in the normal order hierarchy (see Fig.~\ref{fig:mee}), which is much stronger than the previous thought.

Based on the success of our FN model, we also explore its prediction on the neutrino sector, especially, the value of $m_\nu^i$ and $m_{ee}$. We find that the FN mechanism naturally prefers the NO scenario, which is determined by its mathematical structure. Utilizing the neutrino oscillation and cosmology constraint, we calculate the mass spectrum in Fig.~\ref{fig:m3}. It shows that $m_\nu^3$ is almost a constant, i.e. $m_\nu^3\sim 50~\text{meV}$. By adopting this benchmark value we predict that $m_\nu^1$ and $m_\nu^2$ are $\sim1.6$ meV and $\sim 9$ meV respectively, which is perfectly consistent with the current observations. In addition, our model also gives a relatively precise constraint on $m_{ee}$, i.e. $0.503\,{\rm meV} \lesssim m_{ee} \lesssim 18.0\,{\rm meV}$ at $95\%$ C.L. More interestingly, our model can be explored in a near future $0\nu \beta \beta$ experiment, such as LEGEND-1000 \cite{LEGEND:2021bnm}.

In Sec. \ref{sec:fermion} we introduce the factor $\mathcal{N}$ to handle mass hierarchy between up-type quark and down-type quark and the charged lepton (see Eq.~\eqref{eq:yd} and Eq.~\eqref{eq:quark_yukawa}). One possible origin of $\mathcal{N}$ is that there are two different Higgs, like two-Higgs-doublet model~\cite{Branco:2011iw} and supersymmetric theories at high energy. In fact, the FN charge could play the same role as $\mathcal{N}$, e.g. the $\lambda^{c+f}$ in Eq.~\eqref{eq:lambdafactor}. For example, we find that if we reset $U(1)^{d_R}_\text{FN} = \{3.5,2.5,2.5\}$, $U(1)^{e_R}_\text{FN} = \{6,3,2\}$, then $\mathcal{N}$ is not necessary. However, the cost of this is that the discrete gauge symmetry $\mathbf{Z}_{33}$ disappears.

The most distinguished feature of this work is the universality of the coupling $g_{ij}$ (including the $g_{ij}'$ for the neutrino sector). By combining this universality with the FN mechanism, we can successfully explain $15$ SM parameters. Furthermore, its prediction on neutrino mass $m_\nu^i$ is surprisingly consistent with the current observation.
Although the deep meaning behind this universality is unknown, we expect that there may be some hidden principles, and we save further explorations for future works.

\begin{acknowledgements}

T. T. Y. is supported in part by the China Grant for Talent Scientific Start-Up Project and by Natural Science Foundation of China (NSFC) under grant No. 12175134 as well as by World Premier International Research Center Initiative (WPI Initiative), MEXT, Japan. Y. -C. Qiu thanks Kavli IPMU, U. Tokyo for its hospitality, where he finished this work.

\end{acknowledgements}

%\appendix

%\section{FN charge determination}\label{appendix:FN_charges}

\bibliographystyle{utphys}
\bibliography{reference}

\end{document}